\documentclass[10pt,conference,compsocconf]{IEEEtran}
\usepackage{times}

\usepackage{caption}
\usepackage[cmex10]{amsmath}
\usepackage[noadjust]{cite}
\usepackage{lipsum}
\usepackage{algorithmic}
\usepackage{graphicx}
\usepackage{array}
\usepackage{mdwmath}
\usepackage{mdwtab}
\usepackage{eqparbox}
\usepackage{fixltx2e}
\usepackage{stfloats}
\usepackage{siunitx}
\usepackage{url}
\begin{document}
\bstctlcite{IEEEexample:BSTcontrol}
\pagenumbering{gobble}

\title{\textbf{\Large IoT-Based Secure Embedded Scheme for Insulin Pump Data Acquisition and Monitoring}\\[0.2ex]}

\author
{\IEEEauthorblockN{Zeyad A. Al-Odat\IEEEauthorrefmark{1},
Sudarshan K. Srinivasan\IEEEauthorrefmark{1}, 
Eman Al-qtiemat\IEEEauthorrefmark{1}, 
Mohana Asha Latha Dubasi\IEEEauthorrefmark{1}, 
Sana Shuja\IEEEauthorrefmark{2}}
\IEEEauthorblockA{\IEEEauthorrefmark{1}Electrical and Computer Engineering, 
North Dakota State University\\
Fargo, ND, USA\\
\IEEEauthorrefmark{2}Electrical Engineering, COMSATS Institute of Information Technology, \\
Islambad, Pakistan\\
Emails: \IEEEauthorrefmark{1}zeyad.alodat@ndsu.edu,
\IEEEauthorrefmark{1}sudarshan.srinivasan@ndsu.edu,
\IEEEauthorrefmark{1}eman.alqtiemat@ndsu.edu,\\
\IEEEauthorrefmark{1}MohanaAshaLatha.Duba@ndsu.edu,
\IEEEauthorrefmark{2}SanaShuja@comsats.edu.pk
}
}
\maketitle

\begin{abstract}
This paper introduces an Internet of Things (IoT)-based data acquisition and monitoring scheme for insulin pumps. 
The proposed work employs embedded system hardware (Keil LPC1768-board) for data acquisition and monitoring. 
The hardware is used as an abstract layer between the insulin pump and the cloud. 
Diabetes data are secured before they are sent to the cloud for storage. 
Each patient's record is digitally signed using a secure hash algorithm mechanism. 
The proposed work will protect the patient's records from being breached from unauthorized entities, and authenticates them from improper modifications. 
The design is tested and verified using $\mu$Vision studio, the Keil board mentioned above, and an ALARIS 8100 infusion pump. 
Moreover, a test case for a real cloud example is presented with the help of the Center of Computationally Assisted System and Technology. 
This center provided the infrastructure service to test our work.
\end{abstract}

\begin{IEEEkeywords}
IoT; Security; Embedded system.
\end{IEEEkeywords}

\IEEEpeerreviewmaketitle

\section{Introduction}
The physical devices that are linked as Internet of Things (IoT) are continuously increasing~\cite{al2015internet}. 
These devices are allowed to mimic human being's senses.
For example, the use of a smart home as an IoT-based application can turn on the air conditioning system when sensing residents are close \cite{atzori2010internet}\cite{kazi2015iot}. 
The industrial world has moved toward the use of IoT by adding Internet connection to small electronic boards \cite{ungurean2014iot}. 
Moreover, the connections between different primitives are made possible through mobile communications \cite{hinge2013mobile}.

Recent improvements in IoT design helps with the support of some health care systems.
An example is tracking patient's records and biomedical devices using the Internet \cite{catarinucci2015iot}. 
Medical devices for diabetic care have also joined the world of IoT through supporting versatile design options. 
However, security issues need to be addressed to ensure device security and the patient's privacy. 
A system with an authentic security mechanism is required to guarantee the security of patient's records. 
One of the existing methods that can be easily implemented in hardware is the 
Secure Hash Algorithm (SHA)~\cite{harsha2014design}. 
The SHA is an official hash algorithm standard that was standardized by the National Institute of Standards and Technology (NIST). 
SHA is compatible with hardware-level implementation, 
and this makes it one of the more desirable methods for hardware designers to use to secure and authenticate their designs~\cite{pub2012secure}.   

The implementation of IoT technology in hardware has become crucial for high performance. 
The benefit of using hardware to manipulate the increased size of patient's records is that the
hardware allows for high-speed computation to manipulate and retrieve records. 
Therefore, hardware designers have moved toward the use of IoT hardware units in their designs, these units support high-speed computation power for IoT related functions~\cite{boppudi2014data}. 
This paper introduces an IoT-based embedded system for insulin pump data acquisition and monitoring.  
Data related to a patient's diabetes disease along with other health records are stored on the cloud. 
All these data need to be secured and authenticated when they are retrieved from the cloud. 
We use the SHA mechanism in our design to support security and authentication.  

The rest of the paper is organized as follows. 
Section~\ref{sec:section2}  describes related work. 
The proposed methodology is presented in Section~\ref{sec:section3}. 
Results are detailed  in Section~\ref{sec:section4}. 
Section~\ref{sec:section5} concludes the paper.

\section{Related Work}
\label{sec:section2}
There is a lot of recent work in the employment of IoT applications for health monitoring and control~\cite{catarinucci2015iot}\cite{harsha2014design}\cite{rahmani2015smart,hsueh2016next,liu2015secure}. 
A novel IoT-aware smart architecture for automatic monitoring and tracking of the patient, personnel, and biomedical devices, was presented in \cite{catarinucci2015iot}. 
The proposed work built a smart hospital system relying on Radio Frequency Identification (RFID), Wireless Sensor Network (WSN), and smart mobile. 
The three hardware components were incorporated together through a local network, to collect surrounding environment and patient's physiology related parameters. 
The collected data is sent to a control center in real-time, 
which makes all the data available for monitoring and management by the specialist through the Internet.

A smart E-health care system for ubiquitous health monitoring is proposed in~\cite{rahmani2015smart}. 
The proposed work exploits ubiquitous health care gateways to provide a higher-level of services. 
The Gateways are the bridging point between IoT and applications (software or hardware). 
This work studied significant ever-growing demands that have an important influence on health care systems. 
The proposed work suggests an enhanced health care environment where control center burdens are transferred to the gateways. 
The security of this scheme is taken into consideration as the system deals with substantial health care data. 

 A personalized health care scheme for the next generation wellness technology has been proposed in~\cite{hsueh2016next}. 
 The security of patient's records was addressed in case of data storage and retrieval over the cloud. 
The proposed work established a patient-based infrastructure allowing multiple service providers including the patient, service providers, specialists, and researchers to access the stored data. 
Their work was implemented on a cloud-based platform for testing and verification. 
Liu \textit{et al}.~\cite{liu2015secure} have presented a scheme for secure sharing of personal health records in the cloud. 
The health records are ciphered before they are stored in the cloud. 
The proposed work uses Cipher-Text Attribute-Based Signcryption Scheme (CP-ABSC) as an access control mechanism. Using this scheme, they are able to get fine-grained data access over the cloud.  

The use of embedded micro-controllers for data monitoring and acquisition has also been previously explored. 
The Keil LPC1768 micro-controller has been used in two different schemes for medical device control~\cite{harsha2014design}\cite{boppudi2014data}. 
In~\cite{harsha2014design}, an online design of patient's data monitoring system was presented. 
The proposed work employed an Advanced RISC Machine (ARM) Cortex M3 microprocessor embedded on the Keil LPC1768 board. 
Pulse, temperature, and gas sensors were used to collect the patient's medical parameters. 
The LPC1768 board was used as the hardware layer between the Internet and the medical sensors. 
A data acquisition and control system using the ARM Cortex M3 microprocessor was also presented in~\cite{boppudi2014data}. 
Monitored sensor data are sent to the Internet using an Ethernet-controlled interface.
The interface was built using an Keil LPC1768 board. 
The proposed work employed two sensing devices (temperature and accelerator-meter) to collect data from the surrounding environment. 
The collected readings are sent to the Internet through the Ethernet interface. 
According to the uploaded readings, a specialist can change the behavior of the device through an Internet browser.

In the next section, we will present our proposed IoT-based secure data acquisition and monitoring scheme. 
The integration between the embedded architecture and the cloud-computing based storage will be discussed in detail. 

\section{Proposed Methodology}
\label{sec:section3}
We provide some background information before getting into the details of the proposed methodology. 
In the subsequent text, we provide a brief description of the secure hash standard.

\subsection{Background: Secure Hash Algorithm}
SHA takes a message with an arbitrary size and produces a message hash through some calculations. 
The process is defined in equation~(\ref{eq:hash}).

\begin{equation}
h = H(M)
\label{eq:hash}
\end{equation}
where, $M$ is the input message and $h$ is the digest generated using the hash algorithm $H$.

\begin{figure}[htbp]
    \centering
    \includegraphics[width=\columnwidth]{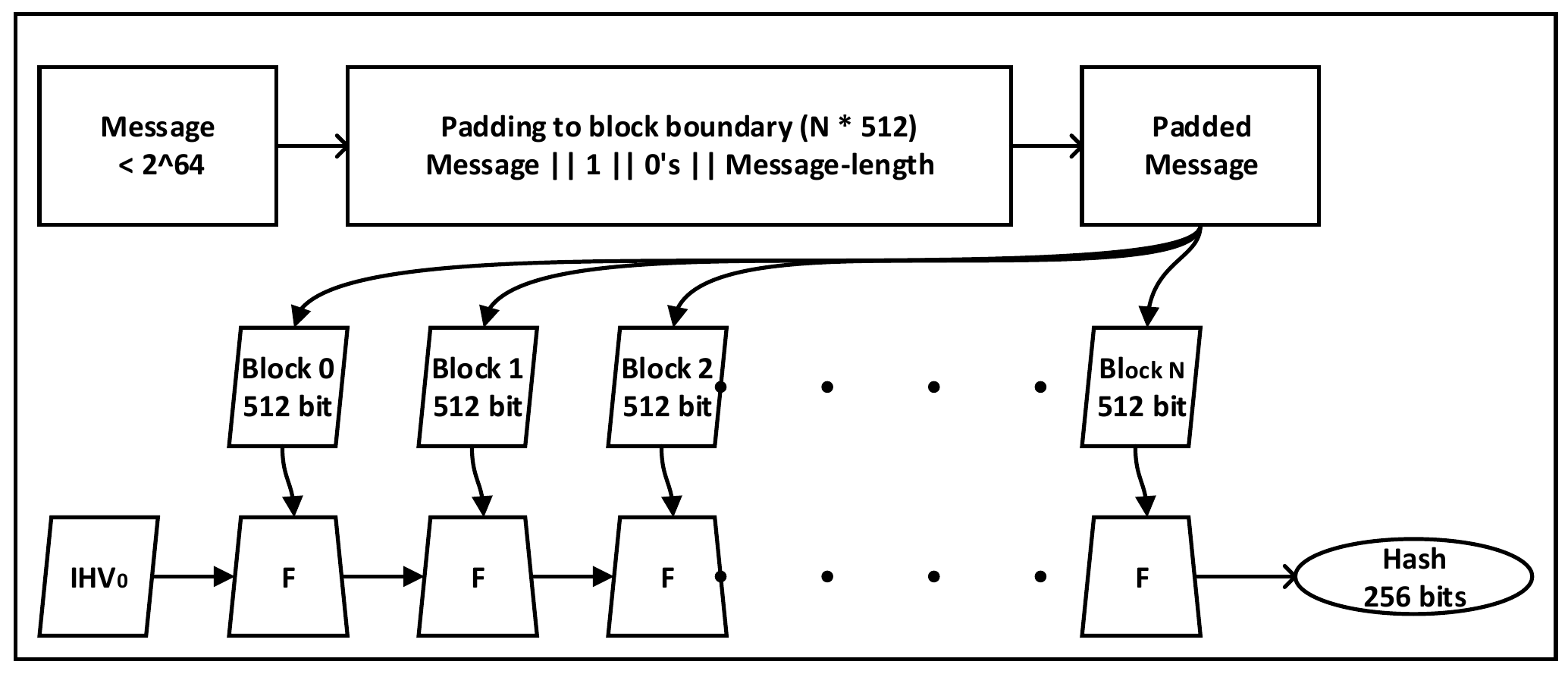}
    \caption{SHA general architecture: Padding, dividing, compression, and computation of SHA-256.}
    \label{fig:sha}
    \vspace{-0.2in}
\end{figure}

In our scheme, the SHA2/256 standard is employed for securing the patient's records. 
Data are signed with the SHA2/256 before they are stored in the cloud. 
The stored records are made available for research centers and medical institutions. 
Figure~\ref{fig:sha} depicts the general procedure that is used to compute the hash for any given message. 
The input message of size less than $2^{64}$ is padded first to make its size a multiple of $512$. 
The full message is divided into equal size blocks of 512-bits. 
The blocks are then processed sequentially using compression function $F$, and Initial Hash Value ($IHV_0$) that are defined in~\cite{pub2012secure}.  

At the end of the process, the hash value that is generated from the last block produces the final $256$ bits hash. 
A detailed description of the secure hash algorithm can be found in~\cite{pub2012secure}.

\subsection{General Architecture of the Proposed Scheme}

\begin{figure}[htbp]
    \centering
    \includegraphics[width=\columnwidth]{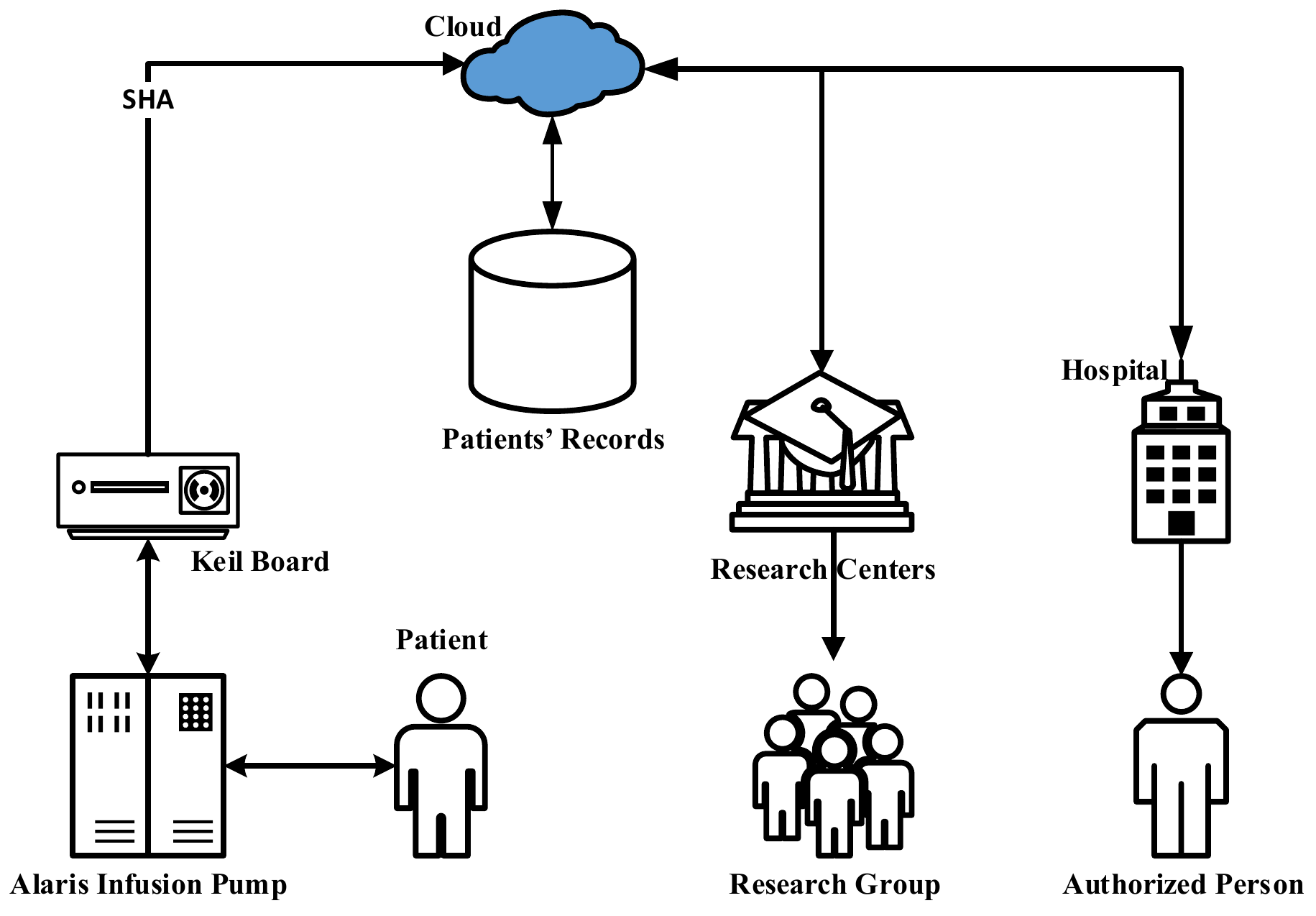}
    \caption{General architecture of the proposed scheme.}
    \label{fig:general}
    \vspace{-0.2in}
\end{figure}

The proposed design for the secure IoT-based embedded system is depicted in Figure~\ref{fig:general}. 
A patient using the Alaris 8100 infusion pump will take preset insulin doses regularly. 
The Alaris infusion pump is controlled and monitored by the Keil Cortex M3 board through a serial connection. 
All dosage related records that are given to the patient are sent to the cloud through the Keil board using an Ethernet connection. 
To secure and authenticate the recorded data, they are digitally signed using the SHA compression function. 
The signature and patient's records are then stored in the cloud.

In the cloud, a Secure Socket Shell (SSH) is provided to entities authorized to access the data. 
For instance, a physician can follow up with a patient's case using a mobile application or a web browser. 
Furthermore, research institutions are given the authorization to access health records upon agreements made between
patient, medical centers, and research institutions.

The security and authenticity of the health care records are verified using the SHA signature. 
The SHA value is computed after the health records or prescription commands are generated. 
Then the generated SHA is appended to the corresponding data (health record or preset control command). 
The health record and its signature are remained correlated in all places (cloud, hospital, and patient's side). 
For instance, the physician in the hospital confirms that the record is received without altering using SHA signature. 
When the health record is received at the hospital, SHA computation will be carried out. 
The resultant SHA will be compared with the appended SHA value. 
Once both SHA values are equal, the record will be affirmed to its corresponding patient. 
Otherwise, the health record will be discarded as is does not belong to the patient. 

\begin{figure}[htbp]
    \centering
        \includegraphics[width=\columnwidth]{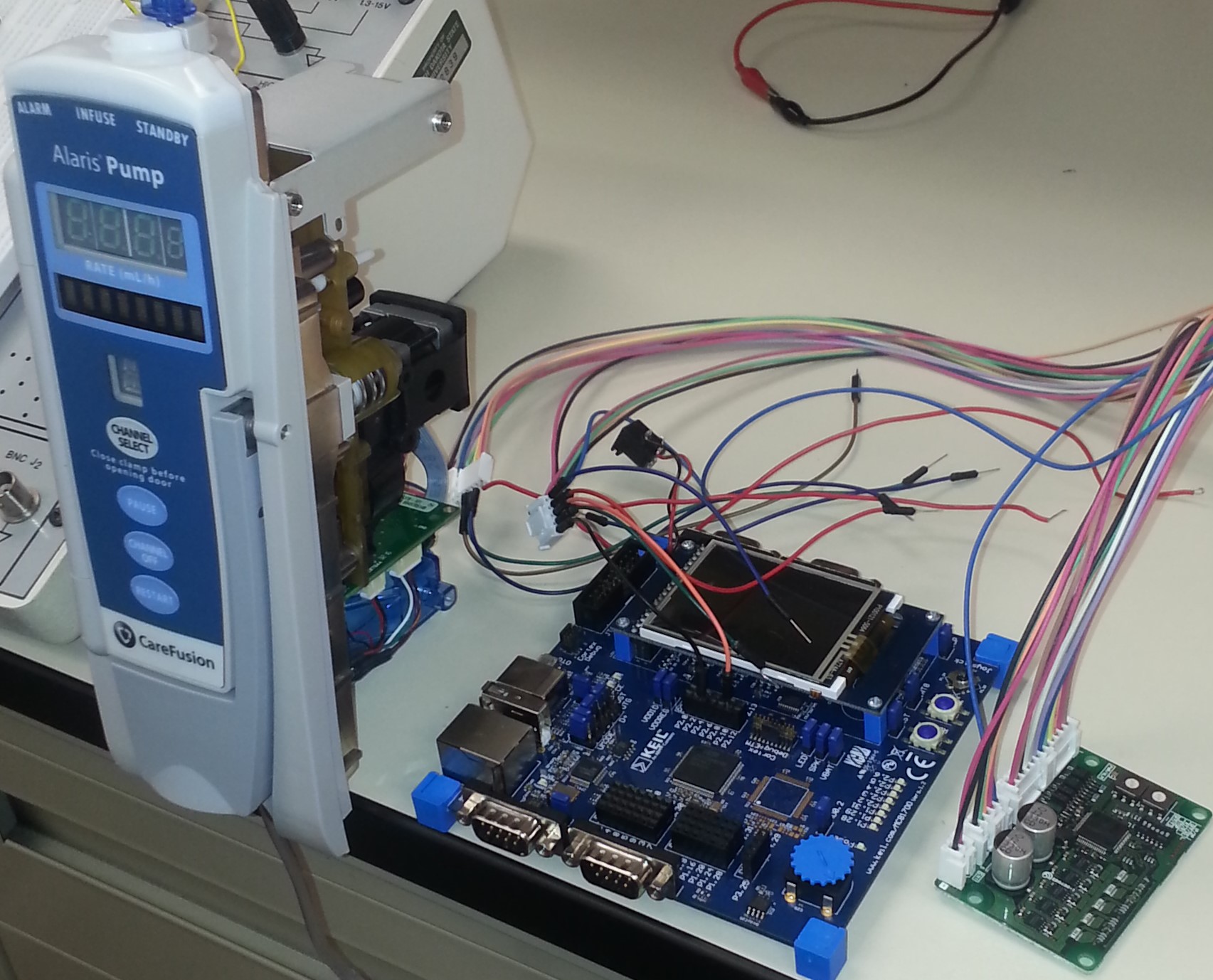}
    \caption{Connection of Alaris Infusion Pump 8100 with Keil 1768 PCB board}
    \label{fig:Circuit}
    \vspace{-0.2in}
\end{figure}

In the case of the preset control command, this command is generated from the hospital and appended with its corresponding hash value.  
The preset control command and the SHA signature are sent through the cloud to the infusion pump. 
At the patient's end, the hardware takes the responsibility to check the genuineness of the received control command by SHA computation and comparison. 
The Keil micro-controller computes the SHA value for the received preset control command and then compares the result with the appended SHA value. 
Once authorized, the preset control command is passed to the infusion pump for a new schedule. 
Figure~\ref{fig:Circuit} shows the connection between the Keil LPC1768 board and the Alaris 8100 infusion pump.  
  
In the case of a fault exception, all Cortex-M processors (including Keil LPC1768) have a fault exception mechanism embedded inside the processor. 
If any fault is detected, the corresponding exception handler will be executed~\cite{alkim2016newhope}.
The hardware setup was done at the NDSU-Electrical and Computer Engineering laboratories. 
The infusion pump was first disassembled, interfaced with the Keil micro-controller, and then programmed using a serial cable and the Keil $\mu$Vision studio~5.

\section{Results and Discussion}
\label{sec:section4}

\begin{figure}[htbp]
\centering
        \includegraphics[scale=0.75]{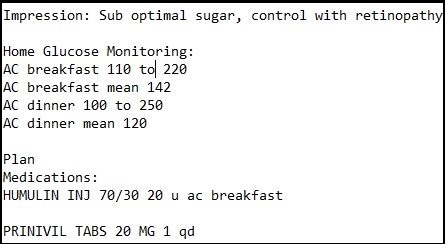} % first figure itself
        \caption{Snipped health record from the original sample}
                \label{fig:sample_orig}
\end{figure}

\begin{figure}[htbp]
\centering
        \includegraphics[scale=0.75]{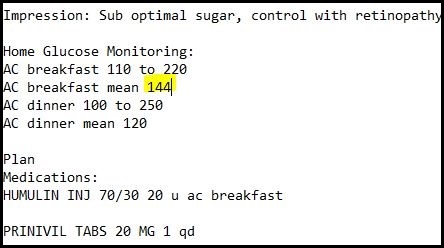} % first figure itself
        \caption{Snipped health record from the modified sample}
                \label{fig:sample_mod}
                \vspace{-0.2in}
\end{figure}

The proposed design has been tested and verified using data from~\cite{BibEntry2018Oct}. 
The sample data includes glucose levels in the patient's body during a 24 hour period, a patient's profile information, and the patient's medical information. 
A snipped portion of the sample data is shown in Figures~\ref{fig:sample_orig}. 
Figure~\ref{fig:sample_mod} shows the modified sample data. 
The data are stored in the cloud regularly, where each copy has its designated SHA value. 
To ensure the integrity of data, SHA-256 is applied to both sides (cloud and patient) after any query from either side.


The record is valid if its generated hash value on the patient's side is the same as the hash value on the cloud side.
Table~\ref{tab:sample} shows a valid record hash. 
However, two different hash values are depicted for the same record in Table~\ref{tab:sample_modified}, because the received record on the patient's side has been altered. Accordingly, the corresponding hash value has also been altered. 
The micro-controller will detect the alteration and discard the received record.

\begin{table}[htbp]
    \centering
    \caption{SHA-256 HASH VALUES OF THE SAMPLE DATA ON BOTH SIDES.}
            \begin{tabular}{ll}
        \textbf{Cloud side}:& \begin{tabular}[c]{@{}l@{}}\texttt{14b93acf-ccdcbe40-ea3795be-c1073498-}\\
                                                                                        \texttt{51a96c90-6cedfc9c-49d8e2cf-a141befb} \\ 
                    \end{tabular}  \\ \hline
        \textbf{Patient side}:& \begin{tabular}[c]{@{}l@{}}\texttt{14b93acf-ccdcbe40-ea3795be-c1073498-}\\
                                                                                            \texttt{51a96c90-6cedfc9c-49d8e2cf-a141befb} 
                \end{tabular} 
        \\ \hline
    \end{tabular}
    \label{tab:sample}
\end{table}
\begin{table}[htbp]
    \centering
        \caption{SHA-256 HASH VALUES OF THE ORIGINAL AND MODIFIED SAMPLE DATA ON BOTH SIDES.}
            \begin{tabular}{ll}
        \textbf{Cloud side}:& \begin{tabular}[c]{@{}l@{}}\texttt{14b93acf-ccdcbe40-ea3795be-c1073498-}\\
                                                                                        \texttt{51a96c90-6cedfc9c-49d8e2cf-a141befb} \\ 
                    \end{tabular}  \\ \hline
        \textbf{Patient side}:& \begin{tabular}[c]{@{}l@{}}\texttt{358c4f29-f0e2bb60-8efa35d4-a88a6b3b-}\\
                                                                                            \texttt{58939ffd-deebf824-8065c195-b834b8cd} 
                \end{tabular} 
        \\ \hline
    \end{tabular}
    \label{tab:sample_modified}
\end{table}

The patient's intervention for the proposed design is limited to turn on and off the infusion pump. However, the future work of our design will upgrade the patient's privileges, e.g., change the infusion pump schedule according to predefined levels.
 
\section{Conclusion and Future work}
\label{sec:section5}

We have presented a secure IoT-based embedded data acquisition and control scheme. 
The work employed three modules: Keil micro-controller, LPC1768 board, and Alaris 8100 infusion pump. 
Secure Hash Algorithm standard SHA-256 is used to ensure the authenticity of the system. 
The authenticity of the proposed work was verified with a cloud storage utility using a real sample record. 
The results show that any altering in the health record is going to be identified immediately, thus the patient remains safe from false prescriptions. 
In future, we plan to apply the proposed scheme to hand-held glucose devices.

\section*{Acknowledgment}
This publication was funded by a grant from the United States Government and the generous support of the American people through the United States Department of State and the United States Agency for International Development (USAID) under the Pakistan - U.S. Science \& Technology Cooperation Program. The contents do not necessarily reflect the views of the United States Government.   
\linebreak
\bibliographystyle{IEEEtran}
\bibliography{bibo}

\end{document}